\documentclass[onecolumn,12pt]{article}
\usepackage{amsfonts,amsmath,amssymb}
\usepackage{xcolor}
\usepackage{graphicx,subcaption}
\usepackage{hyperref}
\hypersetup{pdftex,colorlinks=true,allcolors=blue}
\usepackage[margin=1.4in, left=1.5cm,right=1.5cm, top=2cm]{geometry}
\usepackage{cite}

\newcommand{\RN}[1]{%
  \textup{\uppercase\expandafter{\romannumeral#1}}%
}

\newcommand{\del}{\partial}

\newcommand{\bulk}{\mt{bulk}}

\newcommand{\jnt}{\mt{jnt}}

\newcommand{\be}{\begin{equation}}
\newcommand{\ee}{\end{equation}}
\newcommand{\bea}{\begin{eqnarray}}
\newcommand{\eea}{\end{eqnarray}}
\newcommand{\beq}{\begin{equation}}
\newcommand{\eeq}{\end{equation}}
\newcommand{\beqa}{\begin{eqnarray}}
\newcommand{\eeqa}{\end{eqnarray}}
\newcommand{\beqar}{\begin{eqnarray*}}
\newcommand{\eeqar}{\end{eqnarray*}}

\def\bal#1\eal{\begin{align}#1\end{align}}

\newcommand{\reef}[1]{(\ref{#1})}

\newcommand{\mt}[1]{\textrm{\tiny #1}}

\newcommand{\mC}{\mathcal{C}}

\newcommand{\CA}{{\cal C}_\mt{A}}

\newcommand{\nn}{\nonumber}

\newcommand{\cv}{{\cal C}_\mt{V}}

\newcommand{\ctL}{\ell_\mt{ct}}

\usepackage{xspace}
 
\begin{document}
\begin{titlepage}
\begin{center}
\vskip 2 cm
{\LARGE \bf  Complexity Growth of Dyonic Black holes \\  \vskip 0.25 cm with Quartic Field Strength Corrections }\\
\vskip 1.25 cm
 Hamid Razaghian\footnote{razaghian.hamid@gmail.com}

\vskip 1 cm
{{\it Department of Physics, Faculty of Science, Ferdowsi University of Mashhad\\}{\it P.O. Box 1436, Mashhad, Iran}\\}
\vskip .1 cm
 \end{center}
\begin{abstract}
In this paper, according to CA duality, we study complexity growth of dyonic RN-type black holes with Quartic Field Strength Corrections  ($F^4$ corrections) to the matter action in general $D\geq 4$-dimensions and find the behavior of action growth, similar to the case of the normal RN black holes, is different between electric and magnetic black holes  which is unexpected since violates the electromagnetic duality.
For restoring this duality, we add the Maxwell boundary term (at order 3-derivative) to the action and discuss the outcomes of the addition.
Also, we have used another method that introduces UV finite cut off at AdS boundary to evaluate the complexity growth rate of dyonic black holes with and without $F^4$ corrections.
In this method, without adding a surface term, late time growth rate of complexity exhibits expected behavior.
\end{abstract}
\end{titlepage}


\tableofcontents
\section{Introduction}
Recently, the combination study of holographic entanglement entropy and quantum information shed light on the understanding of quantum gravity \cite{VanRaamsdonk:2010pw}.
However, the entanglement entropy may not be enough to probe the degrees of freedoms in black holes interior since the volume of black holes continues growing even if spacetimes reach thermal equilibrium \cite{Susskind:2014moa}.
Instead, {\it complexity} was proposed to be the correct quantity to characterize the interior of black holes.

``Quantum complexity" is defined by the minimal number of simple gates needed to prepare a target state from a reference state.
From the holographic viewpoint, the quantum complexity of the state in the boundary theory is dual to some bulk gravitational quantities which are called ``holographic complexity".
Complexity was originally conjectured to be proportional to the maximum volume of a codimension-one surface bounded by the CFT slices, 
\begin{align}
\cv=\frac{V}{Gl},
\end{align}
  which is called ``Complexity=Volume" (CV) duality \cite{Susskind:2014rva,Stanford:2014jda,Susskind:2014jwa}.
The length scale $l$ is chosen according to situations.
In order to eliminate the ambiguities in CV duality, ``Complexity=Action" (CA) duality was proposed \cite{Brown, Brown2},  which states that the quantum complexity of a particular  state $\vert \psi(t_L,t_R)\rangle$ on the boundary is given by
\begin{align}
\CA(|\psi(t_L,t_R)\rangle)=\frac{I_{\rm WDW}}{\pi \hbar},\label{eq.02040002}
\end{align}
where $I_{\rm WDW}$ is the on-shell action evaluated on a certain region of spacetimes, called the Wheeler-DeWitt (WDW) patch, which is enclosed by the past and future light sheets sent into the bulk spacetime from the timeslices $t_L$ and $t_R$.
CA duality does not involve any ambiguities encountered in CV duality.
In this paper, we only focus on the CA conjecture. 

The complexity grows linearly with time, and according to the definition, complexity growth rate is the speed of quantum computations.
Considering black holes as the densest memories \cite{tHooft:1993dmi, Susskind:1994vu, Bekenstein:1980jp}, it is conjectured that black holes are the fastest computers in nature.
There exist a bound for the speed of quantum computation. 
Lloyd proposed a bound on the speed of quantum computation in  \cite{Lloyd}.
Brown and collaborators generalized Lloyd’s bound and conjectured that there exists a bound on the growth rate of  complexity at late time, which is
\begin{align}
\frac{d\mC}{dt}\leq \frac{2E}{\pi \hbar}.\label{eq.02040612}
\end{align}
Calculations show that static neutral black holes saturate the bound.
It has been extensively  studied  complexity growth of black holes in different gravity systems to examine CA duality and Lloyd’s bound 
\cite{Moosa:2017yiz,HosseiniMansoori:2017tsm,Mahapatra:2018gig,Chapman:2016hwi,Carmi:2016wjl,Kim:2017lrw,Yang:2016awy,Chapman:2018dem,Chapman:2018lsv,
Moosa:2017yvt,Fan:2018xwf,Bernamonti:2019zyy,Swingle:2017zcd,Alishahiha:2018tep,An:2018xhv,Jiang:2018pfk,Kim:2017qrq,Yang:2019gce,Guo:2019vni,Cai:2016xho,
Lehner:2016vdi,Huang:2016fks,Cai:2017sjv,Cano:2018aqi,Jiang:2018sqj,Jiang:2019fpz,Feng:2018sqm,Alishahiha:2017hwg,
Carmi:2017jqz,Jiang:2019pgc,Liu:2019smx,Flory:2018akz,Flory:2019kah,Ghodrati:2018hss,Ghodrati:2017roz,Zhou:2019jlh,Ghodsi:2020qqb}.

The authors of refs. \cite{Liu:2019smx, Goto:2018iay, Jiang:2019qea} found that action growth rates vanish for purely magnetic black holes in four dimensions, which is unexpected since the expected late time result is
\begin{align}
\frac{dI}{dt}\sim TS,\label{eq.02040627}
\end{align}
Therefore, there is no electric-magnetic duality in their solution.
In order to improve the situation, a Maxwell boundary term
\begin{align}
\int_{\partial M} d\Sigma_\mu {F}^{\mu\nu}A_\nu
\end{align}
were proposed to be included in the action and by this method, the electromagnetism duality was restored.
However, from the point of view of holography, it is of great interest to explore holographic applications of higher derivative on complexity growth.
In this paper, we would like to generalize the discussions of \cite{Goto:2018iay}  and study the action growth of dyonic black holes in Einstein-Maxwell gravity with quartic field strength corrections in the matter action. 
These corrections are subsets of $\alpha'$ corrections for higher dimensional AdS space with a general $U(1)$ and we review black hole solutions and their thermodynamic quantities in the grand canonical ensemble \cite{Anninos:2008sj}.
The purpose of this paper is to examine whether adding boundary term to the bulk action  is needed for restoring electric-magnetic duality in the presence of corrections of higher derivatives Maxwell fields.
Following the same method of \cite{Goto:2018iay}, we conclude that one must add the same Maxwell boundary terms (but in higher derivatives)  to prevent vanishing of action growth rates for purely magnetic black holes.

Another approach to restoring electric-magnetic duality of dyonic black holes in action growth rate is proposed in \cite{Akhavan:2018wla, Alishahiha:2018swh} by introducing finite boundary cut off $r=r_c$ and  the complexity growth rate is calculated at the finite cut off geometry. 
In this method, in order to have a consistency behavior of growth rate complexity with Lloyd's bound at late time, one is forced to introduce a new cut off surface $r=r_0$  behind the outer horizon whose value is fixed by the boundary cut off $r_c$ \cite{Akhavan:2018wla}.
The other important thing in this method is that the boundary term is not needed and  variational principle does not change.

Similar calculations are done in  \cite{Alishahiha:2018swh, Alishahiha:2019cib} and \cite{Hashemi:2019xeq} but for  Jackiw-Teitelboim (JT) gravity, (near extremal) black branes and electrical charged black holes respectively.
In this paper, we have also used this approach, and have studied the growth rate of holographic complexity of $F^2$ and $F^4$ black holes by introducing finite boundary cut off $r=r_c$ and cut off surface $r=r_0$.
By finding a relation between $r_0$ and $r_c$, and substitute $r_0$ in terms of $r_c$ and inner and outer horizons $r_\pm$ in action growth, we observe that growth rate of complexity is electromagnetic-dual.

The paper is organized as follows.
In section \eqref{sec.0204}, we obtain a solution of dyonic black holes in Einstein-Maxwell gravity with correction at fourth-order derivative in electromagnetic field strength and then calculate some basic thermodynamic quantities of this system.
In section \eqref{sec.0304}, we evaluate the complexity growth by considering all contributions of bulk, surface, joint, and counterterm actions.
First, we investigate the CA conjecture without the Maxwell boundary term, and then we examine action growth by adding the boundary term of matter field and discuss the outcomes of the addition.
In section \eqref{sec.Cutoff}, we will compute holographic complexity for $F^2$ and $F^4$ theory in the presence of a UV radial cut off $r_c$ and a behind horizon cut off $r_0$ and show that consistency with Lloyd's bound at the late time enforces us to find a relation between $r_0$ and $r_c$ and this relation will restore electric-magnetic duality of the growth rate of complexity.
We summarize our calculations in the last section.

\section{Geometry of $F^4$ Theory}\label{sec.0204}
We begin our discussion by studying following bulk action in $ D $ dimensions spacetime  (we set $G_N=1$):
\begin{align}
I_\mt{bulk} = & \frac{1}{16\pi}\int_\mathcal{M} d^{D} x \sqrt{-g}\left(\mathcal{L}_{\rm EH}+ \mathcal{L}_{\rm Max}+  \mathcal{L}_{\rm F^4}\right)\label{eq.111559}
\end{align}
where
\begin{align}
\mathcal{L}_{\rm EH}&=R -2 \Lambda ;\qquad \Lambda=-\frac{(D-1)(D-2)}{2L^2}\nn\\
\mathcal{L}_{\rm Max}&=-   F_{\mu \nu} F^{\mu \nu};\qquad F_{\mu\nu}=\partial_\mu A_\nu-\partial_\nu A_\mu \nonumber\\
\mathcal{L}_{F^4}=& \alpha' \left[B_1 ( F_{\mu \nu} F^{\mu \nu})^2+B_2 F_{\mu \nu} F^{\nu \alpha} F_{\alpha \beta} F^{\beta \mu}\right].\label{eq.14091541}
\end{align}
In above,  $A$ is the gauge field, $F$ is the electromagnetic field strength,  $L$  is the AdS radius  and $B_1,B_2$ are arbitrary constants.
From the  point of view of string theory, the $F^4$ term is of order $\alpha'$.
Here, the parameter $\alpha'$ is a small dimensionless expansion parameter that controls the effects of quartic  electromagnetic field strength corrections.

Now by taking variation of the bulk action with respect to metric $g_{\mu\nu}$, one obtains the field equation:  
\begin{align}
E_{\mu\nu}=&\;R_{\mu \nu }  -2 F_{\mu }{}^{\alpha } F_{\nu \alpha } +  4\alpha'\Big(B_1  F_{\alpha\beta} F^{\alpha\beta } F_{\mu }{}^{\gamma } F_{\nu \gamma }  +   B_2 F_{\alpha }{}^{\gamma } F_{\beta \gamma } F_{\mu }{}^{\alpha } F_{\nu }{}^{\beta }\Big)\nn\\
&-  \frac{1}{2}\Big(R-2\Lambda - F_{\alpha \beta } F^{\alpha \beta }+  \alpha' B_1   F_{\alpha \beta } F^{\alpha \beta } F_{\gamma \delta } F^{\gamma \delta }+\alpha' B_2   F_{\alpha }{}^{\gamma } F^{\alpha \beta } F_{\beta }{}^{\delta } F_{\gamma \delta } \Big)g_{\mu \nu }  \label{eq.111415}\\
=&\;0\nn
\end{align}
The modified form of the Maxwell equations in a curved background is:
\begin{align}
\frac{\partial I_{\rm bulk}}{\partial A^{\mu}}=\nabla_\alpha \widetilde{F}_\mu{}^\alpha=0,\label{eq.111744}
\end{align}
where
\begin{align}
\widetilde{F}_{\mu\nu}&=-2 \frac{\partial }{\partial F^{\alpha\beta}}\left(\mathcal{L}_{\rm Max}+\mathcal{L}_{F^4}\right)\nn\\
&=4\left( F_{\alpha\beta}-2\alpha'\left(B_1 F_{\alpha\beta} F_{\gamma\delta}F^{\gamma\delta}+B_2 F_{\alpha}{}^\gamma F_\beta{}^\delta F_{\gamma\delta}\right) \right).
\label{eq.17081910}
\end{align}
We solve equation of motions for static AdS planar black holes in $D=2n+2$ dimensions as
\begin{align}
&ds^2=-f(r)dt^2+\frac{1}{f(r)}dr^2+r^2 (dx_1^2+dx^2_2+\cdots +dx^2_{2n-1}+dx^2_{2n}),\nn\\
& F=E(r) dr\wedge dt+\left( \Psi_1(x) dx_1\wedge dx_2 +\Psi_2(x) dx_3\wedge dx_4+ \cdots+\Psi_n(x) dx_{2n-1}\wedge dx_{2n}\right).\label{eq.171659}
\end{align}
Bianchi identity 
\begin{align}
\nabla_{[r}F_{x_i x_{i+1}]}=0,
\end{align}
indicates that $\Psi_i(x)$'s are independent of $r$.
The $x_i$-component of eq. \eqref{eq.111744}, has a term that is $r-$dependent and must be equal to some constant:
\begin{align}
-4 \left(r^4+4\alpha' B_1 E(r)^2 r^4 \right)\partial_{x_j} F_{x_i}{}^{x_j}=c.
\end{align}
If $\Psi_i$'s are $x$-dependent functions, then  behavior of $E(r)$  as an electric field will be incorrect at $r\to \infty$:
\begin{align}
E(r)^2=\frac{c}{r^4}-1, \quad \lim_{r\to\infty} E(r)^2=-1
\end{align}
So, $\Psi_i(x)$'s are constant functions and we set them as follows:
\begin{align}
\Psi_i(x)=p,
\end{align}
where $p$ is an integration constant and will soon be related with the magnetic charge.


The $r$-component of eq. \eqref{eq.111744} is
\begin{align}
&\left((D-2) E(r)+r \frac{\partial E(r)}{\partial r} \right) 
-\frac{2\alpha' B_1(D-2) p^2}{r^4}\left((D-6)E(r)+r\frac{\partial}{\partial r} E(r)\right)+\nn\\
& \qquad 2\alpha'(2B_1+B_2)E(r)^2\left((D-2) E(r)+3 r \frac{\partial E(r)}{\partial r}\right)=0.
\end{align}
The above equation is exactly solvable.
Physically,  when $B_1$ and $ B_2 $ goes to zero, the expected solution must leads to the usual Coulomb’s law,  and it should not diverges if $r$ approaches infinity.
So the general solution  is as follows:
\begin{align}
E(r)=\frac{1}{6^{1/3}}\left(
\frac{1}{\Delta}\left(1-\frac{2(D-2)\alpha' B_1 p^2}{ r^4}\right)-\frac{\Delta}{6^{1/3}\alpha'(2B_1+B_2)}
\right)
\end{align}
where
\begin{footnotesize}
\begin{align}
\Delta(r)=&\frac{1}{r^{D+4}}\Big[\Big(-9\alpha'^2(2B_1+B_2)^2q r^{2(7+D)}\nn\\
&+\sqrt{3\alpha'^3 (2B_1+B_2)^3r^{4(3+D)}\left(27\alpha' (2B_1+B_2)q^2  r^{16}+2 r^{2D}(r^4-2\alpha'B_1 (D-2)p^2)^3\right)}\Big)^{1/3}\Big]
\end{align}
\end{footnotesize}

\noindent
and $q$ is an integration constant which we will soon related  it with the electric charge.
To solve $A_t=\Phi_e(r)$, we have to simplify $E(r)$ as much as we can.
This difficulty is due to considering the effects of   quartic field strength corrections on the matter action.
We can expand $E(r)$ in terms of $\alpha'$ corrections up to $\mathcal{O}(\alpha'^2)$:
\begin{align}
E(r)=\frac{q }{r^{D-2}}\left(1-\frac{2\alpha' (2B_1+B_2)q^2  }{r^{2D-4}}+\frac{2\alpha' B_1(D-2)p^2}{r^4}  \right)+\mathcal{O}(\alpha'^2). \label{eq.111309}
\end{align}
As  can be seen, the leading order is Coulomb's law, and if $\alpha'\to 0$, it's all that survives.
Next-to-leading order terms are corrections of Coulomb's law and Reissner-Nordstrom solutions  by $\alpha'$ parameter.

The electric scalar potential field is integration of $E(r)$:
\begin{align}
\Phi_e(r)=\int^{\infty}_{r} E(r)\;dr=\frac{q}{(D-3)r^{D-3}}-\frac{2\alpha' (2B_1+B_2)q^3}{(3D-7)r^{3D-7}}+\frac{2\alpha'B_1(D-2)qp^2}{(D+1)r^{D+1}}.\label{eq.01301423}
\end{align}
The constant integration in $\Phi_e(r)$ is chosen such that  $\Phi_e(\infty)=0$.
In $r=r_+$, the $\Phi_e(r_+)$ is chemical potential associated to the outer horizon, while in $r=r_-$, $\Phi_e(r_-)$ does not have a boundary dual.

The $ rr $-component of equation of motions in \eqref{eq.111415} is (up to   $O(\alpha')$):
\begin{align}
&\frac{1}{2}(D-2)r f'(r)+\frac{1}{2}(D-2)(D-3)f(r)+\Lambda r^2+\frac{q^2}{r^{2D-6}}+\frac{(D-2)p^2}{2r^2}\nn\\
&-\alpha'\left(\frac{(2B_1+B_2)q^4}{r^{4D-10}}+\frac{(D-2)((D-2)B_1+B_2)p^4}{2r^6}-\frac{2B_1(D-2)(q^2p^2)}{r^{2D-2}}\right)=0.
\end{align}

The solution of above equation is as follows:
\begin{align}
f(r)=&-\frac{\mu}{r^{D-3}}-\frac{2\Lambda r^2}{(D-1)(D-2)}+\frac{2q^2}{(D-2)(D-3)r^{2D-6}}-\frac{p^2}{(D-5)r^2}+\nn\\
&\alpha'\left(-\frac{2(2B_1+B_2)q^4}{(3D-7)(D-2)r^{4D-10}}+\frac{((D-2)B_1+B_2)p^4}{(D-9)r^6}+\frac{4B_1q^2p^2}{(D+1)r^{2D-2}}\right),\label{eq.02011636}
\end{align}
and $\mu$ is an integration constant.
In this paper, we only consider solutions that have two RN-type horizons which are located at $r=r_\pm$.
As can be seen, only in $D=4$, the function $f(r)$ is symmetric under change of $q$ and $p$.

Total electric and magnetic charges of black hole can be calculated by using Gauss's law.
The total electric charge is
\begin{align}
Q_e&=\frac{1}{4\pi} \int_{r\to \infty} d x_1  dx_2\ldots dx_{2n} (\star F)_{x_1x_2\ldots x_{2n}}=\frac{1}{4\pi} \int_{r\to \infty} d x_1  dx_2\ldots dx_{2n} \sqrt{-g}F_{rt}\nn\\
&=\frac{1}{4\pi} \int_{r\to \infty}  d x_1  dx_2\ldots dx_{2n}\;(r^{D-2}) E(r)=\frac{q}{4\pi}\omega_2^n \label{eq.1010}
\end{align}
Similarly, the total magnetic charge is
\begin{align}
Q_m=\frac{1}{4\pi} \int_{r\to \infty} dx_i  dx_j F_{x_i x_j}=\frac{D-2}{8\pi} \int_{r\to \infty} dx_1  dx_2 \;(p ) =\frac{(D-2)p\omega_2}{8\pi} \label{eq.102030}
\end{align}

The total mass of black hole is (by ADM Approach \cite{Abbott:1981ff})
\begin{align}
M=\frac{n\omega_2^n \mu}{8\pi}
\end{align}
where $\mu$ is determined by evaluating the metric function on horizon $(f(r=r_+)=0)$:
\begin{align}
\mu=&-\frac{2\Lambda r^{D-1}_+}{(D-1)(D-2)}+\frac{2q^2}{(D-2)(D-3)r^{D-3}_+}-\frac{p^2 r^{D-5}_+}{D-5}\nn\\
&-\alpha'\left(\frac{2(2B_1+B_2)q^4}{(3D-7)(D-2)r_+^{3D-7}}-\frac{((D-2)B_1+B_2)p^4 r^{D-9}_+}{D-9}-\frac{4B_1q^2p^2}{(D+1)r^{D+1}_+} \right)
\end{align}
The magnetic scalar potential can be calculated by the first law of black hole thermodynamics:
\begin{align}
dM=TdS+\Phi_e dQ_e+\Phi_m dQ_m,\label{eq.01301425}
\end{align}
where entropy $S$ and temperature $T$ are:
\begin{align}
S=\frac{\omega_2^n}{4}r^{2n}_+\;,\qquad T=\frac{1}{4\pi} f'(r)\Big\vert_{r=r_+}.
\end{align}
So, the magnetic scalar  potential is
\begin{align}
\Phi_m(r)&=\left(\frac{\partial M}{\partial Q_m}\right)_{S,Q_e}\nn\\
&=-\frac{p\,\omega^{n-1}}{(D-5)}r^{D-5}+\alpha'\omega^{n-1}\left(
\frac{2((D-2)B_1+B_2)p^3  }{(D-9)r^{9-D}}+\frac{4B_1 q^2q_m  }{(D+1)r^{D+1}}\right).\label{eq.01301436}
\end{align}
As can be seen, in $D=4$ dimension, if we change electric and magnetic charges ($q_e\leftrightarrow q_m$), the electric and magnetic potentials are also interchanged ($\Phi_e\leftrightarrow \Phi_m$).
But in higher  dimensions ($D>4$), this symmetry does not occur.
\section{Evaluating the Growth Rate of Action}\label{sec.0304}
\subsection{Without Maxwell Boundary Term}
\begin{figure}[!t]
    \centering
    \begin{subfigure}[b]{0.4\textwidth}
        \centering
        \includegraphics[height=4in]{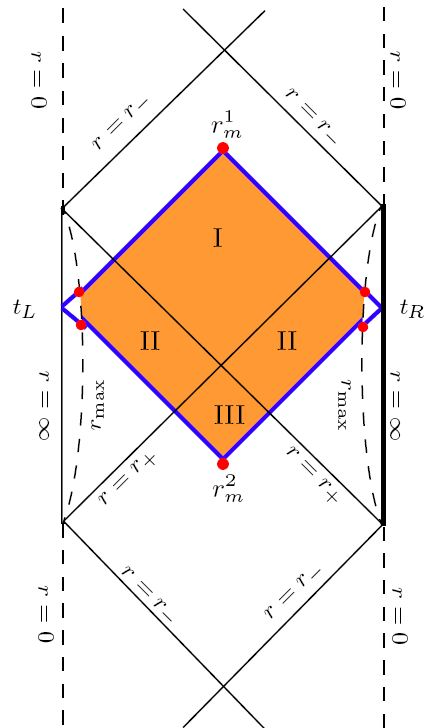}
       \caption{ }\label{fig1a}
    \end{subfigure}
    \begin{subfigure}[b]{0.4\textwidth}
\centering
        \includegraphics[height=4in]{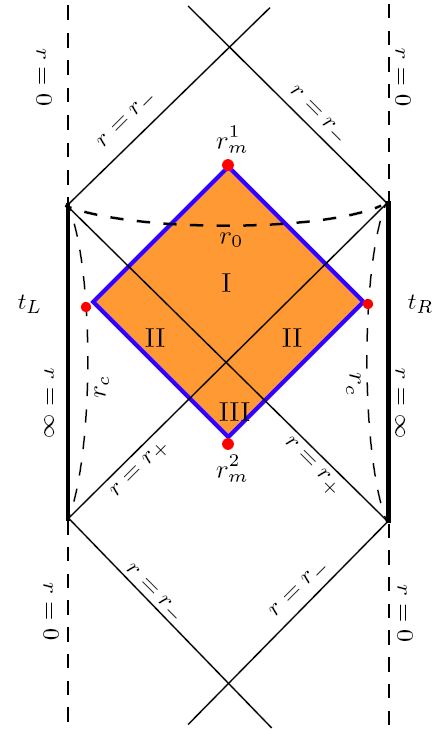}
       \caption{ }\label{fig1b}
    \end{subfigure}
\caption{\textbf{(a)}. The causal structure of a charged asymptotically AdS black hole which has two horizons located at $r_+$ and $r_-$.
The WDW patch is divided into three regions, I, II, and III.
At later times, the future meeting point, $r^1_m$, approaches the inner horizon $r_-$, and the past meeting point, $r^2_m$, approaches the outer horizon $r_+$.\\
\textbf{(b)}.
Penrose diagram of the Dyonic AdS black hole with a radial boundary finite cut off $r_c$ and also $r=r_0$ cut-off behind the outer horizon.
}
\label{fig.1}
\end{figure}

To calculate the holographic complexity, we have to compute on-shell action on the WDW patch.
The WDW patch is a spacetime area enclosed by null sheets.
Figure \ref{fig.1}(a) illustrate the change of the WDW patch of this type of black hole with two RN-type horizons in the Penrose diagram.
To regulate the divergence near the AdS boundary, we have introduced a cut-off surface $r=r_\infty$.
As we will see, action growth of this type of black hole is formally identical with AdS-RN black holes, and correction to field strength does not change the causal structure of the spacetime, however, an electromagnetic field affects the growth of action through the correction in higher orders.
The WDW patch is general non-smooth, so we employ the method proposed in \cite{Goto:2018iay,NullBound} to calculate the action.

The time evolution of action is  determined by the evolution of the future ($r_m^1$)  and  past ($r_m^2$)  meeting points of the light rays  bounding the WDW patch, which is characterized by
\begin{equation} 
t_L+ r^*(\infty) - r^*(r_m^{1})=0,
\qquad
t_R-r^*(\infty)+r^*(r_m^{2})=0,\label{eq.02091212}
\end{equation}
where $r_-\leq r^1_m,r^2_m\leq r_+$ and $r^*(r)$ is tortoise coordinate which is defined as $r^*(r)=\int^\infty_r dr/f(r)$.
So in our convention, $r^*(\infty)=0$.
Here $t_L\;(t_R)$, is time coordinate of left (right) boundary located at the cut-off surface  $r=r_{\rm max}$ and the boundary time is $t=t_L+t_R$.
For later convenience, we always choose $t_R=t_L=t/2$.

As time elapse, the meeting points $r_m^1$ and $r_m^2$ begin to grow with time as follows ($f(r^{1,2}_m)<0$, because position is in black holes interior)
\begin{equation} 
\frac{d r_m^{1}}{d t} = \frac{f(r_m^{1})}{2}\leq 0,
\qquad
\frac{d r_m^{2}}{d t} = - \frac{f(r_m^{2})}{2}\geq 0.\label{eq.02082243}
\end{equation}
Thus, as  time grows,  the $r^2_m$   increases from the inner horizon $r_-$ to the outer horizon $r_+$ while the  $r^1_m$ behaves precisely in the opposite way \cite{Fan:2019aoj}.

The on-shell gravitational action is composed of different parts:
\begin{equation}
I_\mt{tot} = I_\mt{bulk} + I_\mt{surf} + I_{\mt{ct}}  \, ,\label{eq.171507}
\end{equation}
where $I_\mt{bulk}$ is bulk action in \eqref{eq.111559} and the next term $I_\mt{surf}$ contains various surface terms that we have to add to make the variational principle well-defined for the metric,
\begin{align}
I_\mt{surf}=&  \frac{1}{8\pi } \Big(
\Sigma_{T_i}\int_{\partial\mathcal{M}_{T_i}} d^{D-1} x \sqrt{|h|} K 
+\Sigma_{S_i}\mathrm{sign}(S_i)\int_{\partial\mathcal{M}_{S_i}} d^{D-1} x \sqrt{|h|} K+ \nn\\
&    \Sigma_{j_i}\mathrm{sign}(j_i) \oint d^{D-2}x \sqrt{\sigma} \eta_{j_i}+  \Sigma_{N_i}\mathrm{sign}(N_i)\int_{\partial\mathcal{M}_{N_i}} d\lambda\, d^{D-2} \theta \sqrt{\gamma} \kappa\nn\\
& + \Sigma_{m_i}\mathrm{sign}(m_i) \oint d^{D-2} x \sqrt{\sigma} a_{m_i}\Big) \, ,\label{eq.171510}
\end{align}
where $S_i$, $T_i$ and $N_i$ labels spacelike, timelike and null boundary respectively.
The first and second terms are the Gibbons-Hawking-York (GHY) surface terms \cite{York, GH} for smooth timelike and spacelike segments of the boundary respectively (in terms of the trace of the extrinsic curvature $K$).
$\eta_{j_i}$  is the Hayward joint term  \cite{Hay1, Hay2}, which appears at the intersection of two non-null boundary segments.
Forth and fifth contribution are the surface and joint terms for null boundary segments \cite{NullBound}. 
The parameter $\lambda$, is the parameter of the null generator $k$ on the null segments, and $\kappa$ measures the failure of $\lambda$ to be an affine parameter on the null generators which is derived from $k^a\nabla_a k^b=\kappa k^b$.
The $a_{m_i}$ is the joint term at the intersection of these null boundary segments with any other boundary segment.
The signature $\mathrm{sign}(N_i)$, $\mathrm{sign}(j_i)$ and $\mathrm{sign}(m_i)$ are determined through the requirement that the gravitational action is additive \cite{NullBound}.

The null surface counterterm $ I_{\mt{ct}}$ in total action is not necessary for the action principle, but its contribution   is due  to ensure reparametrization invariance on the null boundaries \cite{NullBound}, and is:
\begin{align}\label{Counterterm}
I_\mt{ct}= \frac{1}{8 \pi } \int_{\mathcal{B}'} d \lambda \, d^{D-2} \theta \sqrt{\gamma}\, \Theta \,\log \left(\ctL \Theta \right) ,
\end{align}
where $\gamma $ is the determinant of induced metric, $\Theta$ is the scalar expansion defined by $\Theta =\nabla_a k^a= \del_\lambda \log{\sqrt \gamma}$ and $\ctL$ is an undetermined length scale.

Now, for evaluating the action \reef{eq.171507}, we will consider nonvanishing contributions of all parts separately.
\subsubsection*{Bulk contribution}
First, we evaluate the time derivative of the bulk term in eq.~\eqref{eq.111559}.
By using  the dyonic geometry \eqref{eq.02011636}   and the Maxwell field \reef{eq.111309}, the integrand in the bulk action is given by
\begin{align}
\mathcal{L}(r)&= R -2\Lambda  -   F_{\mu \nu} F^{\mu \nu}+\alpha'\left(B_1 ( F_{\mu \nu} F^{\mu \nu})^2+B_2 F_{\mu \nu} F^{\nu \alpha} F_{\alpha \beta} F^{\beta \mu}\right) \nn\\
&=  -\frac{4\Lambda}{D-2}+\frac{4q^2}{(D-2)r^{2D-4}}-\frac{2p^2}{r^4}+\nn\\
&\quad \alpha'\left(-\frac{4(2B_1+B_2)q^4}{(D-2)r^{4D-8}}+\frac{6((D-2)B_1+B_2)p^4}{r^8}-\frac{8B_1q^2p^2}{r^{2D}}\right)
 \label{BulkIntegrand}
\end{align}
We then write the bulk action as
\begin{equation}
I_\mt{\bulk} =\frac{\omega^n_2}{16\pi } \,\int  dr\, r^{D-2}\, \mathcal{L}(r) \ \int  dt \label{eq.101213}
\end{equation}
where $\omega_2=\int dx_1 dx_2=\int dx_3 dx_4=\cdots=\int dx_{2n-1}dx_{2n}$.

We split WDW patch into three pieces (see fig. \ref{fig.1}(a)) by following \cite{Cano:2018aqi,Carmi:2017jqz}):
\begin{align}
I^{\rm I}_{\rm bulk} &=\frac{2\omega^n_2}{16\pi } \,\int^{r_+}_{r^1_m}  dr\, r^{D-2}\, \mathcal{L}(r) (\frac{t}{2}-r^*(r)) ,\nn\\
I^{\rm II}_{\rm bulk} &= \frac{4\omega^n_2}{16\pi } \,\int^{\infty}_{r_+}  dr\, r^{D-2}\,  \mathcal{L}(r)(-r^*(r) ) ,\nn\\
I^{\rm III}_{\rm bulk} &=\frac{2\omega^n_2}{16\pi } \,\int^{r_+}_{r^2_m}  dr\, r^{D-2}\, \mathcal{L}(r) (-\frac{t}{2}-r^*(r)). \label{eq.02091206}
\end{align}

With differentiating of $I_\mt{\bulk}$ with respect to $t$, we find :
\begin{align}
\frac{d I_{\bulk}}{d t} &=
\frac{\omega^n_2}{16\pi } \int_{r_m^{1}}^{r_+} r^{D-2} \mathcal{L}(r) dr-\frac{\omega^n_2}{16\pi } \int_{r_m^{2}}^{r_+} r^{D-2} \mathcal{L}(r) dr  
= \frac{\omega^n_2}{16\pi } \int_{r_m^{1}}^{r_m^{2}} r^{D-2} \mathcal{L}(r) dr \nn\\
&= \frac{\omega^n_2 }{4\pi }
\Bigg[ 
\frac{\Lambda r^{D-1}}{(D-1)(D-2)}-\frac{q^2}{(D-2)(D-3)r^{D-3}}-\frac{p^2r^{D-5}}{2(D-5)}\nn\\
&\quad+\alpha'\left(\frac{(2B_1+B_2)q^4}{(3D-7)(D-2)r^{3D-7}} +\frac{3((D-2)B_1+B_2)p^4}{2(D-9)r^{9-D}}+\frac{2B_1 q^2p^2}{(D+1)r^{D+1}}\right)
\Bigg]
 \Bigg|^{r_m^{2}}_{r_m^{1}},\label{BulktDer}
\end{align}
where we have adopted the relation  \eqref{eq.02091212} for the meeting points.

We can rewrite above expression in terms of  electric and magnetic charge and potentials as  follows (eq. \eqref{eq.01301423}, \eqref{eq.1010}, \eqref{eq.102030} and \eqref{eq.01301436}):
\begin{align}
\frac{d I_{\bulk}}{d t} &=\Bigg[\frac{\omega^n_2 }{4\pi }\frac{\Lambda r^{D-1}}{(D-1)(D-2)}-\frac{Q_e\Phi_e}{D-2}+\frac{Q_m\Phi_m}{D-2}\nn\\
&+64\alpha'\pi^3\omega^n_2\Big(-\frac{ (2B_1+B_2)Q_e^4}{(3D-7)(D-2)r^{3D-7}\omega^{4n}_2}
+ \frac{8((D-2)B_1+B_2)Q_m^4 }{(D-9)(D-2)^4r^{9-D}\omega^{4}_2}\nn\\
&+  \frac{8B_1   Q_e^2 Q_m^2}{(D+1)(D-2)^2r^{D+1}\omega_2^{2n+2}}\Big)\Bigg]
 \Bigg|^{r_m^{2}}_{r_m^{1}}\label{eq.103029}
\end{align}
\subsubsection*{Surface contributions}
The WDW patch is cut off by two UV regulator surfaces at  large timelike hypersurface $\phi=r=r_\mt{max}$  where $r_{\rm max}$ is the UV cutoff (see fig \ref{fig.1}(a).)
The (future-directed) unit normal to any surface $  r = $constant  inside the future horizon is given by\footnote{$\epsilon$ is defined as
\[
n^\mu n_\mu=\epsilon=\begin{cases}
-1 & \text{if}\ \Sigma\ \text{is spacelike}\\
+1 & \text{if}\ \Sigma\ \text{is timelike}
\end{cases}
\]
}
\begin{align}
n_\mu=\epsilon\frac{1}{\sqrt{g^{\alpha\beta}\partial_\alpha\phi\partial_\beta\phi}}\partial_\mu\phi=\frac{1}{\sqrt{f(r)}}\partial_\mu r
\end{align}
The trace of the extrinsic curvature is given by
\begin{align}
K&=\nabla_\mu n^\mu=\frac{1}{r^{D-2}}\frac{\partial}{\partial r}\left(r^{D-2} \sqrt{f(r)}\right)=\frac{1}{2\sqrt{f(r)}} \left(\del_r f(r) + \frac{2(D-2)}{r} f(r)\right)\,,
\end{align}
and the volume element is 
\begin{equation} 
d\Sigma=  d^{D-1} x\sqrt{|h|}=(d\omega_{2})^n\sqrt{f(r)}r^{D-2}dt 
\end{equation} 
So as a result, we obtain
\begin{equation}
I_{\text{surf}}^{\text{cutoff}}=\frac{	r^{D-2} (\omega_{2})^n}{4\pi } \left(\partial_r f(r)+\frac{2(D-2)}{r}f(r)\right)
\left(r^*_{\infty}-r^*(r)\right)
\biggr{|}_{r=r_{\max}}.
\label{harvey}
\end{equation}
Since the surface contribution $I_{\text{surf}}^{\text{cutoff}}$ is independent of time, its time rate is zero and we will ignore them  here.

Further, without loss of generality, with affinely-parametrized null normals, all $\kappa $ surface terms on null boundaries  in eq.~\eqref{eq.171510} vanishes. 
\subsubsection*{Joint contributions}
There are joint terms in action \eqref{eq.171510} where two of the boundary surfaces intersect.
From the fig. \ref{fig.1}(a), it is easy to see that the Hayward term $\eta$ vanish because all of the joints involve at least one null surface, so we only consider the last term in action  \eqref{eq.171510}
\beq
I_{\jnt}= \frac1{8\pi }\int_{\Sigma'} d^{D-2}x \sqrt{\sigma}\, a \, ,
\label{ActJ2}
\eeq
where $a$ is standard corner term for Einstein's gravity which defined as \cite{NullBound},
\begin{equation}
a= \begin{cases}
 \epsilon \log{ |k \cdot t| } \qquad\text{for  spacelike-null joint with }\epsilon = -\mbox{sign}(k \cdot t)\, \mbox{sign}(k \cdot \hat s)\,, \\
\epsilon \log{| k \cdot s |}\qquad \text{for  timelike-null joint with }\epsilon = -\mbox{sign}(k \cdot s) \, \mbox{sign}(k \cdot \hat t)\,, \\
		 \log|k\cdot \bar k/2| \qquad \text{for a null intersecting surface}\,.
\end{cases}\label{ActJ2a}
\end{equation}
and $\hat s$ and $\hat t$ ($\hat k$) are auxiliary unit {\it vectors} (null {\it vector}) in the tangent space of the spacelike/timelike boundary surface, which are orthogonal to the junction and point outwards from the boundary region of interest.

We have four timelike-null joints where the null boundary meets the cut-off surface $r=r_{\mt{max}}$.
In this case ${\bf \hat t}=\hat t^\mu\,\del_\mu = \del_t $  and   $\epsilon=-1$.
Unit normal vector  to this surface is
\[
s=s_\mu dx^\mu=\frac{dr}{\sqrt{f(r_{\max})}}
\]
and null normal is 
\[
\textbf{k}=dv\vert_{r=r_{\max}}=\left(dt+\frac{dr}{f(r)}\right)\Bigg\vert_{r=r_{\max}}.
\]
Implicitly, we have normalized this null normal at the asymptotic AdS boundary such that $\textbf{k}\cdot \hat{\textbf{t}}=1$.
The corresponding contribution (including the   factor of 4) to this joint is
\begin{equation}\label{CornerCut}
I_{\jnt,\text{cut}}= \frac{\omega_{2}^n}{4 \pi }\,  r^{D-2}\log{f  (r)}\,\Big|_{r=r_{\mt{max}}} \,.
\end{equation}
As such as eq.~\eqref{harvey}, the null joint contributions at the UV cutoff surface have no time dependence and its time rate is zero, too.

The next step is evaluating the joint contributions coming from the meeting points, $r=r_m^1$, and $r_m^2$.
We use the following outward-directed null normal vectors:
\beq
{\rm Right}\,:\quad{\bf k}_\mt{R} = -\xi dt +\xi \frac{dr}{f(r)}\,;\qquad
{\rm Left}\,:\quad{\bf k}_\mt{L} = \xi dt + \xi\frac{dr}{f(r)}\,.
\label{rapido}
\eeq
where $\xi$ is a normalization constant, i.e., $k\cdot\partial_t\mid_{r\to\infty}=\pm\xi$.
With these choices, we have  ${\bf k}_\mt{R} \cdot {\bf k}_\mt{L} = 2\xi^2/f$, so that  
\begin{equation} 
a = -\log\biggl( \frac{|f|}{\xi^2} \biggr)\,. 
\end{equation} 
The joint terms can then be evaluated as
\begin{align}\label{RNjoints1}
I_{\text{joint}} &=\frac{1}{8\pi }\int (d\omega_{2})^n r^{D-2} a\nn\\
&=- \frac{\omega^n_2}{8\pi } \, \left[ (r_m^1)^{D-2} \log \left[ \frac{|f  (r_m^1) |}{\xi^2} \right] +  (r_m^2)^{D-2} \log \left[ \frac{|f  (r_m^2) |}{\xi^2} \right]  \right] \, .
\end{align}
Meeting points $r_m^1$ and $r_m^2$ are time dependent as determined by eq.~\reef{eq.02082243}, so above action has contribution in time rate of change of holographic complexity.
But this contribution is sensitive to the ambiguity through its dependence on the normalization constant $\xi$.

 Using eq.~\eqref{eq.02082243}, the time derivative of eq.~\eqref{RNjoints1} becomes
\begin{align}
\frac{d I_{\text{joint}}}{d t}= \frac{\omega^n_2}{16\pi } \left[ (D-2) r^{D-3} f  (r) \log \frac{|f  (r)|}{\xi^2}  + r^{D-2} \partial_r f  (r) \right]^{r_m^2}_{r_m^1} \, . \label{JnttDer1}
\end{align}
Note that at late times, $r^{1,2}_m$ approach the horizons and so the first term on the r.h.s vanishes because $f(r_\pm)=0$. Hence only the second term contributes to the late-time growth rate:
\begin{align}
\lim_{t\to \infty}\frac{d I_{\text{joint}}}{d t}=& \frac{\omega^n_2}{16\pi }  r^{D-2} \partial_r f  (r) \Big\vert^{r_+}_{r_-}  \label{JnttDer2}\\
=& \Bigg[-\frac{\omega_2^{n}}{4\pi}\frac{\Lambda r^{D-1}}{(D-1)(D-2)}-\frac{(D-3)Q_e\Phi_e}{D-2}-\frac{Q_m\Phi_m}{D-2}\nn\\
&+64\alpha'\pi^3\omega^n_2\Big(\frac{ (2B_1+B_2)Q_e^4}{(3D-7)(D-2)r^{3D-7}\omega^{4n}_2}
-\frac{8((D-2)B_1+B_2)Q_m^4 }{(D-9)(D-2)^4r^{9-D}\omega^{4}_2}\nn\\
&-  \frac{8B_1   Q_e^2 Q_m^2}{(D+1)(D-2)^2r^{D+1}\omega_2^{2n+2}}\Big)\Bigg]
 \Bigg|^{r_m^{2}}_{r_m^{1}}\label{eq.101213}
\end{align}
Note that the $\alpha'$ contribution of eq. \eqref{eq.101213} will cancel  $\alpha'$ contribution of bulk action \eqref{eq.103029}. 
\subsubsection*{Counterterm contribution}
The counterterm for each null surface is given by 
 \begin{equation}\label{Counterterm11}
I_\mt{ct} = \frac{1}{8 \pi } \int_{\mathcal{B}'} d \lambda \, d^{D-2} \theta \sqrt{\gamma}\, \Theta \,\log \left(\ctL \Theta \right) \, ,
\end{equation}
where $\gamma$ is determinant of the induced metric on the meeting points and $\lambda$ is the affine null coordinate on the null segments.
With affine parametrization 
\begin{align}
\lambda =
\frac{r}{\xi},
\end{align}
the expansion takes the form
\begin{align}
\Theta &= \frac{1}{\sqrt{\gamma}}\partial_\lambda \sqrt{\gamma}=\frac{\xi}{r^{D-2}}\partial_r(r^{D-2})=\frac{(D-2)\xi}{r}
\end{align}
So, the counterterm \eqref{Counterterm11}   becomes
\begin{align}
 I_\mt{ct} &=  \frac{(D-2) \omega_{2}^n}{4\pi }\Big( \int^{r_{\max}}_{r^1_m} \!\!\!\!\!
d r \, r^{D-2}  \log \frac{(D-2)\xi\ctL}{r}
+ \int^{r_{max}}_{r^2_{m}} \!\!\!\!\!
d r \, r^{D-2}  \log \frac{(D-2)\xi\ctL}{r}\Big)
\nonumber\\
&= \frac{r_{\text{max}}^{D-2}\omega^n_2}{2\pi(D-2)}\left[ (D-2)\log \left(\frac{(D-2) \xi  \ctL }{r_{\text{max}} } \right) + 1\right] \nn\\
&\qquad- \frac{( r_{m}^{1})^{D-2}\omega^n_2}{4\pi(D-2)}  \, \left[ (D-2)\log \left( \frac{ (D-2) \xi   \ctL }{( r_{m}^{1}) } \right) + 1 \right] \nn\\
&\qquad- \frac{( r_{m}^{2})^{D-2}\omega^n_2}{4\pi(D-2)}  \, \left[ (D-2)\log \left( \frac{ (D-2) \xi   \ctL }{( r_{m}^{2}) } \right) + 1 \right].\label{TotalCT1} 
\end{align}
Time dependence only comes from terms evaluated at the meeting points on the second and third lines.
The time derivative of eq.~\eqref{TotalCT1} is
\begin{equation}\label{CTtDer1}
\frac{d I_{\text{ct}} }{d t} =\left[  \frac{(D-2)r^{D-3} f (r)\omega^n_2}{8\pi} \,  \log \left( \frac{(D-2) \xi  \ctL  }{r } \right) \right]^{r_m^2}_{r_m^1} \, .
\end{equation}
At late times, $r^{1}_m\to r_{- }$ and $r^2\to r_+ $, so $f(r_{\pm})=0$ and therefore the above term vanishes:
\begin{align}
\lim_{t\to \infty}\frac{d I_{\text{ct}}}{d t}= 0.\label{eq.0100}
\end{align}
\subsubsection*{Total growth rate}
The growth rate of the holographic complexity \reef{eq.02040002} at late time is then given by the sum of eqs.~\eqref{eq.103029} and  \eqref{eq.101213}  which yields
\begin{align}
 \lim_{t \to \infty}\frac{d \mathcal{C}_A}{d t}   
&=-\frac{1}{\pi} Q_e\Phi_e(r)\Big\vert^{r_+}_{r_-}, \label{square}
\end{align}
where $\Phi_e(r)$ is electric scalar potential \eqref{eq.01301423}.

Similar to 4-dimensional dyonic black holes in Einstein-Maxwell gravity, only the electric charge and electric potential appear in $\delta I/\delta t$.
It seems that, regardless of the order of derivative of electromagnetic theory, only the electric charge affects the late-time growth rate of the black holes action.
Because the magnetic potential does not appear in action growth,  if we consider a purely magnetically charged black hole with $Q_e = 0$, the action growth rate always vanishes! 
This result violates the electromagnetic duality of the Maxwell theory.

\noindent\textbf{Note.} Unlike  4-dimensional dyonic black holes in Einstein-Maxwell gravity, the magnetic charge affects the action growth of the dyonic black hole through some constant in $\Phi_e$ (see eq. \eqref{eq.01301423}). 

\subsection{Adding Maxwell Boundary Term}\label{sec.0404}
 In order to restore the late-time result $\frac{dI}{dt}\sim TS$ for purely magnetic black holes and to solve abandon of electric-magnetic duality in the growth rate of complexity, we try to add some appropriate boundary terms of Maxwell field in action \eqref{eq.171507} (proposed in \cite{Goto:2018iay}).

In \cite{Liu:2019smx}, a second-order formalism proposed that yields a general formula for required boundary terms to restoring the electromagnetic duality.
In this formalism, the boundary term is
\begin{align}
I_{\mu Q}=\frac{\gamma}{16\pi  }\int_{\partial M} d\Sigma_\mu \widetilde{F}^{\mu\nu}A_\nu,\qquad \widetilde{F}_{\mu\nu}=-2\frac{\partial \mathcal{L}_F(F_{\mu\nu})}{\partial F_{\mu\nu}},
\end{align}
where $\gamma$ is a free parameter.
If we set $\mathcal{L}_{F}$ as follows
\begin{align}
\mathcal{L}_{F}(F_{\mu\nu})&=\mathcal{L}_{\rm Max} +\mathcal{L}_{F^4}\nn\\
&=-   F_{\mu \nu} F^{\mu \nu}+\alpha' \left[B_1 ( F_{\mu \nu} F^{\mu \nu})^2+B_2 F_{\mu \nu} F^{\nu \alpha} F_{\alpha \beta} F^{\beta \mu}\right],
\end{align}
then $\widetilde{F}_{\mu\nu}$ is equation \eqref{eq.17081910}:
\begin{align}
\widetilde{F}^{\mu\nu}=4\left( F^{\mu\nu}-2\alpha'\left(B_1 F^{\mu\nu} F_{\gamma\delta}F^{\gamma\delta}+B_2 F^{\mu\gamma} F^{\nu\delta} F_{\gamma\delta}\right) \right)\label{eq.17081910}
\end{align}
So the relevant boundary term is
\begin{align}
I_{\mu Q}=\frac{\gamma}{4\pi  }\int d \Sigma_\mu 
\left(F^{\mu\nu}-2\alpha'\left(B_1 F^{\mu\nu} F_{\gamma\delta}F^{\gamma\delta}+B_2 F^{\mu\gamma} F^{\nu\delta} F_{\gamma\delta}\right) \right)A_\nu. \label{eq.02040849}
\end{align}
Unlike Gibbons-Hawking surface term, this boundary term was not required by the variational principle, so does not affect the field equations.
But introduce different boundary conditions of the electromagnetic fields \cite{Goto:2018iay}.
Actually, after adding the boundary term \eqref{eq.02040849}, the Dirichlet boundary condition $\delta A_a=0$ requirement becomes  Neumann boundary condition $n^\mu\partial_\mu \delta A_a=0$ for  $\gamma=1$ and mixed boundary conditions for general  $\gamma$.

By using Stokes's theorem and equation of motion $\nabla_\mu \tilde{F}^{\mu\nu}=0$, we can write Maxwell boundary term as an on-shell bulk integration
\begin{align}
I_{\mu Q}\Big\vert_{\rm on-shell}=\frac{\gamma}{8\pi  }\int d^D x\sqrt{-g}\left(F_{\mu\nu}F^{\mu\nu}-2\alpha'(B_1 ( F_{\mu \nu} F^{\mu \nu})^2+B_2 F_{\mu \nu} F^{\nu \alpha} F_{\alpha \beta} F^{\beta \mu})\right)
\end{align}
The effect of  this term on integrand of bulk action \eqref{eq.111559} is
\begin{align}
&\mathcal{L}(r)_{\mu Q} =-4 \gamma   \Bigg( \frac{  q^2}{ r^{2D-4}}-\frac{ (D-2) )p^2}{2r^4}\nn\\
&  +\alpha'\left( \frac{ (2B_1+B_2)   q^4}{r^{4D-8}}-\frac{ ((D-2)B_1+B_2) (D-2)  p^4}{r^8}\right)
\Bigg),\label{BulkIntegrand2}
\end{align}
and in eq. \eqref{eq.101213}  is
\begin{align}
\frac{dI_{\mu Q} }{dt}=\gamma(Q_e \Phi_e -Q_m\Phi_m)\Big\vert^{r^+}_{r^-}.
\end{align}
So, the growth rate of the holographic complexity at late time is (at order $O(\alpha')$)
\begin{align}
 \lim_{t \to \infty}\frac{d \mathcal{C}_A}{d t} =&\frac{1}{\pi}\left[\frac{dI_{\rm bulk}}{dt}+\frac{dI_{\rm joint}}{dt}+\frac{dI_{\mu Q}}{dt}\right]\nn\\
=& -\frac{1}{\pi}\left[(1-\gamma)Q_e \Phi_e+\gamma Q_m \Phi_m\right]\Big\vert^{r^+}_{r^-}.\label{eq.1707}
\end{align}
This result is similar with the four-dimensional AdS-RN black holes \cite{Goto:2018iay}.
So by choosing $\gamma=\frac{1}{2}$, under interchanging $Q_e\leftrightarrow Q_m$ and $\Phi_e\leftrightarrow \Phi_m$, the growth rate of complexity is invariant, which it agrees with electric-magnetic duality in four dimensions.
If we take  $\gamma=1$, contrary to the  $\gamma=0$ case, only  $Q_m \Phi_m$ contributes to action growth, i.e., action growth rates vanish at late times for purely electrically charged black holes in this case.
In this case, electric charge affects the action growth of the dyonic black hole only through some constant in $\Phi_m$ (see eq. \eqref{eq.01301436}) .

\section{Dyonic Black-holes Theory at Finite Cut Off}\label{sec.Cutoff}
The second resolution for the problem of retaining electromagnetic duality of dyonic black holes is considering a finite boundary cut off $r=r_c$ and also a new cut off surface  $r=r_0$ between inner and outer horizon, which they are shown in figure.\ref{fig.1}(b). 
We observe that the value of cut off $r=r_0$ is related to finite boundary cut off $r=r_c$. 
This relation is a conclusion of consistent Lloyd's bound with late-time behavior of growth rate complexity \cite{Akhavan:2018wla}.
Another important point in this method is that the Maxwell boundary term is no longer needed here and the variational principle does not change.
Similar calculations are done in  \cite{Alishahiha:2018swh, Alishahiha:2019cib} and \cite{Hashemi:2019xeq} but for  Jackiw-Teitelboim (JT) gravity, (near extremal) black branes and electrical charged black holes respectively.

In this method, first, we calculate the complexity growth rate at a finite cut off geometry using the CA proposal. 
Next, we find the energy-momentum tensor at finite radial cut off for the $F^4$ theory and use it to calculate the quasi-local energy. 
By having these two results and using Lloyd's bound, we can find the relation between boundary cut off $r_c$ and behind the outer horizon cut off $r_0$.
If we substitute $r_0$ in terms of $r_c$  in action growth, we observe that the growth rate of complexity is electromagnetic-dual.

\subsection{Action Growth at Finite Cut Off}
Introducing a cut off inside the horizon restricts our access to all regions on the WDW patch located behind the horizon.
This cut off removes the joint point $r_m^1$ from the WDW patch and instead, we would have a space-like boundary at $r=r_0$.
So space-time solutions and evaluation of on-shell action are the same as have been studied in sections \eqref{sec.0204} and \eqref{sec.0304}, but for a new WDW patch that has no joint point $r_m^1$, as shown in the figure \ref{fig.1}(b).

The on-shell gravitational action is:
\begin{equation}
I_\mt{tot} = I_\mt{bulk} + I_\mt{surf} + I_{\mt{ct}}  \, ,\label{eq.0022}
\end{equation}
where $I_\mt{bulk}$ is bulk action in \eqref{eq.111559} and surface action  $I_\mt{surf}$ is composed of:
\begin{align}
I_\mt{surf}&=I_{\rm GHY}+I_{\rm joint}\nn\\
&= - \frac{1}{8\pi }\int_{r=r_0} d^{D-1} x \sqrt{|h|} K  +    \frac{1}{8\pi } \oint_{r=r^2_m} d^{D-2} x \sqrt{\sigma} a   \, ,\label{eq.0025}
\end{align}
where extrinsic curvature $K$ and joint term $a$ are:
\begin{align} 
&K=\frac{1}{2\sqrt{f(r)}} \left(\del_r f(r) + \frac{2(D-2)}{r} f(r)\right)\,,\nn\\
&a = -\log\biggl( \frac{|f|}{\xi^2} \biggr)\,.\label{eq.1722}
\end{align} 
As discussed in section \ref{sec.0304},  at late times the growth rate of counterterm action $I_{\rm ct}$ vanishes  (see eq. \eqref{CTtDer1} and \eqref{eq.0100} ), so we do not need this term in the following.

As before, to calculate the contribution from the bulk action, we split WDW patch into three regions: I, II and III (see figure \ref{fig.1}(b)):
\begin{align}
I^{\rm I}_{\rm bulk} &=\frac{2\omega^n_2}{16\pi } \,\int^{r_+}_{r_0}  dr\, r^{D-2}\, \mathcal{L}(r) (\frac{t}{2}-r^*(r)) ,\nn\\
I^{\rm II}_{\rm bulk} &= \frac{4\omega^n_2}{16\pi } \,\int^{r_c}_{r_+}  dr\, r^{D-2}\,  (-r^*(r) ) ,\nn\\
I^{\rm III}_{\rm bulk} &=\frac{2\omega^n_2}{16\pi } \,\int^{r_+}_{r^2_m}  dr\, r^{D-2}\, \mathcal{L}(r) (-\frac{t}{2}-r^*(r)). \label{eq.1353}
\end{align}
where $ \mathcal{L}(r)$ is the integrand in the bulk action (see  \eqref{BulkIntegrand}).
Note that region I is between $r_0$ and the outer horizon $r_+$.

By differentiating of $I_\mt{\bulk}$ with respect to $t$, we find:
\begin{align}
\frac{d I_{\bulk}}{d t} &=
\frac{\omega^n_2}{16\pi } \int_{r_0}^{r_+} r^{D-2} \mathcal{L}(r) dr-\frac{\omega^n_2}{16\pi } \int_{r_m^{2}}^{r_+} r^{D-2} \mathcal{L}(r) dr  
= \frac{\omega^n_2}{16\pi } \int_{r_0}^{r_m^{2}} r^{D-2} \mathcal{L}(r) dr \nn\\
&= \frac{\omega^n_2 }{4\pi }
\Bigg[ 
\frac{\Lambda r^{D-1}}{(D-1)(D-2)}-\frac{q^2}{(D-2)(D-3)r^{D-3}}-\frac{p^2r^{D-5}}{2(D-5)}\nn\\
&\quad+\alpha'\left(\frac{(2B_1+B_2)q^4}{(3D-7)(D-2)r^{3D-7}} +\frac{3((D-2)B_1+B_2)p^4}{2(D-9)r^{9-D}}+\frac{2B_1 q^2p^2}{(D+1)r^{D+1}}\right)
\Bigg]
 \Bigg|^{r_m^{2}}_{r_0}.\label{BulktDer1}
\end{align}
The GHY term at the spacelike cut off surface  $r=r_0$ is
\begin{align}
I_{\rm GHY}=-\frac{	r^{D-2} (\omega_{2})^n}{16\pi } \left(\partial_r f(r)+\frac{2(D-2)}{r}f(r)\right)
\left(t-2r^*(r)\right)
\biggr{|}_{r=r_0}.
\end{align}
So, the time derivative of $I_{\rm GHY}$ is
\begin{align}
\frac{dI_{\rm GHY}}{dt}\Big\vert_{r=r_0}=-\frac{\omega_{2}^n}{16\pi }	r^{D-2} \left(\partial_r f(r)+\frac{2(D-2)}{r}f(r)\right)
\biggr{|}_{r=r_0}.\label{eq.1709}
\end{align}

As discussed in section \ref{sec.0304}, the spacelike-null joint contributions at the cut-off surface $r=r_0$  have no time dependence and its time rate is zero (see eq.\eqref{CornerCut}).
The only time dependent joint term contribution is at $r=r^2_m$ and it is given by (see eq. \eqref{RNjoints1})
\begin{align}
I_{\rm joint}=-\frac{\omega_2^n}{8\pi}  (r_m^2)^{D-2} \log \left[ \frac{|f  (r_m^2) |}{\xi^2} \right].\label{eq.1732}
\end{align}
Using the relation \eqref{eq.02082243},  the time derivative of eq. \eqref{eq.1732} is
\begin{align}
\frac{d I_{\text{joint}}}{d t}= \frac{\omega^n_2}{16\pi } \left[ (D-2) r^{D-3} f  (r) \log \frac{|f  (r)|}{\xi^2}  + r^{D-2} \partial_r f  (r) \right]\Bigg\vert_{r=r_m^2} \, . \label{1735}
\end{align}
At late times, $r^{2}_m\to r_+$ and so the first term on the r.h.s vanishes because $f(r_+)=0$. Therefore, only the second term contributes to the late-time growth rate:
\begin{align}
\lim_{t\to \infty}\frac{d I_{\text{joint}}}{d t}=& \frac{\omega^n_2}{16\pi }  r^{D-2} \partial_r f  (r) \Big\vert_{r=r_+} \label{1743}
\end{align}
In the following, we will solve the problem in \textbf{four-dimensions}.

By adding eq.\eqref{BulktDer1}, \eqref{eq.1709} and \eqref{1743}, the following equation is obtained:
\begin{align}
\lim_{t\to\infty}\frac{d\CA}{dt}=-\frac{r_0}{\pi}-\frac{q^2_e}{\pi r_+}-\frac{q^2_m}{\pi r_0}-\frac{r^3_0}{L^2\pi}+\frac{\omega}{\pi}+\alpha'\left(\frac{(2B_1+B_2)(-q^4_e+q^4_m)}{5\pi r_0^5}+\frac{2(2B_1+B_2)q_e^4}{5\pi r_+^5}-\frac{4 B_1 q^2_e q^2_m}{5\pi r_+^5}\right).\label{eq.1718}
\end{align}
Now the aim is to compare the above result with Lloyd's bound, which we must  read from the Quasi-local energy spectrum at finite cut off surfce $r=r_c$.
In the following subsection, we calculate quasi-local energy at finite cut-off.
\subsection{Quasi-local energy at finite cut off}
A standard method to find quasi-local energy is Brown-York Hamilton-Jacobi prescription  \cite{Hartman:2018tkw}.
We have a finite cut-off timelike surface boundary at $r=r_c$.
Boundary stress-tensor is derivative of the action with respect to the induced metric on the timelike boundary and Hamiltonian density at the finite cut off $r=r_c$ is the normal projection of  $T_{ij}$ on a co-dimension two surface ($r,t=$cte).

We consider following Euclidean metric at cut off surface $r=r_c$  
\begin{align}
ds^2=h_{ij}dx^i dx^j=f(r)d\tau^2+r^2 d\Sigma^2_{D-2},
\end{align}
and total renormalized action is
\begin{align}
I_{\rm ren}=I_{\rm bulk} +I_{\rm GHY}+I_{\rm ct},\label{eq.1529}
\end{align}
where bulk action $I_{\rm bulk}$ is eq. \eqref{eq.111559},  the Gibbons-Hawking-York (GHY) surface  term is 
\begin{align}
I_{\rm GHY} =   \frac{1}{8\pi }\int_{r=r_c} d^{D-1} x \sqrt{|h|} K ,
\end{align}
and $I_{\rm ct}$ is standard boundary counterterms given by
\begin{align}
I_{\rm ct}=\frac{1}{16\pi G}\int_{r=r_c} d^{D-1}x \sqrt{|h|}\left(
\frac{2(D-2)}{L}+\frac{L}{(D-3)}\mathcal{R}+\frac{L^3}{(D-3)^2(D-5)}\left(\mathcal{R}_{ij}^2-\frac{D-1}{4(D-2)}\mathcal{R}^2\right)+\cdots\right).\label{eq.1557}
\end{align}
In the above relations, tensors $K$ and $\mathcal{R}, \mathcal{R}_{ij}$ are  extrinsic curvature and intrinsic curvature tensors of induced metric  $h_{ij}$  respectively.

The boundary stress-tensor is
\begin{align}
T_{ij}=\frac{2}{\sqrt{|h|}}\frac{\delta I_{\rm ren}}{\delta h^{ij}}=\frac{2}{\sqrt{|h|}}\frac{\delta (I_{\rm bulk}+I_{\rm GHY})}{\delta h^{ij}}+\frac{2}{\sqrt{|h|}}\frac{\delta I_{\rm ct}}{\delta h^{ij}}.\label{eq.1705}
\end{align}
The variation of bulk and GHY action respect to induced metric $h_{ij}$ is
\begin{align}
\delta( I_{\rm bulk}+I_{\rm GHY})=\frac{1}{16\pi }\int_{r=r_c} d^{D-1}x\sqrt{|h|}(K_{ij}-h_{ij}K)\delta h^{ij},\label{eq.1714}
\end{align}
and variation of counterterm action respect to induced metric $h_{ij}$ is
\begin{align}
\delta I_{\rm ct}=&-\frac{1}{16\pi}\int_{r=r_c} d^{D-1}x\sqrt{|h|}\Bigg[\frac{(D-2)}{L} h_{ij}-\frac{L}{D-3}\left(\mathcal{R}_{ij}-\frac{1}{2}h_{ij}\mathcal{R}\right)\nn\\
&+\frac{L^3}{2(D-3)^2(D-5)}h_{ij}\left(\mathcal{R}_{ij}\mathcal{R}^{ij}+\frac{1}{D-2}\nabla_{i}\nabla^i \mathcal{R}-\frac{D-1}{4(D-2)}\mathcal{R}^2\right)\nn\\
&+\frac{L^3}{2(D-3)^2(D-5)} \left(-4\mathcal{R}^{kl}\mathcal{R}_{ikjl}-2\nabla_k\nabla^k \mathcal{R}_{ij}+\frac{D-3}{D-2}\nabla_i\nabla_j \mathcal{R}+\frac{D-1}{D-2}\mathcal{R}_{ij}\mathcal{R}\right)
\Bigg]\delta h^{ij}\label{eq.1715}
\end{align}

 The energy surface density  $\varepsilon$ is defined by the normal projection of  $T_{ij}$ on a co-dimension two surface ($r,t=$cte),
\begin{align}
\varepsilon=T_{\tau\tau}g^{\tau\tau},
\end{align}
and the total quasi local energy (the energy at finite radial cutoff  $r_c$) is given by 
\begin{align}
E_{\rm bulk}=\int_{r=r_c} d^{D-2}x \sqrt{\sigma} \varepsilon, \label{eq.1727}
\end{align}
and $\sigma_{ab}dx^adx^b=r^2_cd\Sigma_{D-2}$.

Using equations \eqref{eq.1705}-\eqref{eq.1715}, one can see that 
\begin{align}
\varepsilon=-\frac{2(D-2)}{16\pi G  r}\sqrt{f(r)}+\frac{(D-2)}{16\pi G}\left(\frac{-2}{L}\right)
\end{align}
so the quasi local energy  \eqref{eq.1727} at finite cut off  $r=r_c$ is
\begin{align}
E_{\rm bulk}=\frac{(D-2)\omega^n_2 r^{D-2}_c}{8\pi G}\left[\frac{1}{L}-\frac{1}{r_c}\sqrt{f(r_c)}\right].
\end{align}
This energy, is related to the energy in boundary field theory by \cite{Hartman:2018tkw}
\begin{align}
E_{\rm bulk}=\frac{L}{r_c}E_{bdry}.
\end{align}
Therefore, the field theory energy is
\begin{align}
E_{\rm bdry}=\frac{(D-2)\omega^n_2 r^{D-1}_c}{8\pi L G}\left[\frac{1}{L}-\frac{1}{r_c}\sqrt{f(r_c)}\right]\label{eq.1914}
\end{align}
\subsection{Complexity at finite cut off}
For small charged black holes, $r_-<r_0<r_+\ll L$,   the late time growth rate of complexity is
\begin{align}
\lim_{t\to \infty}\frac{dC}{dt}=\frac{2}{\pi}\left((E_{\rm bdry}-E_{\rm global})-\Phi_e(r_+)Q_e-\Phi_m(r_+)Q_m\right)\label{eq.1915}
\end{align}
where gravitational quasi-local energy of global AdS energy  $E_{\rm global}$ can be obtained from \eqref{eq.1914}  by $\omega, q, p=0$.

Substituting back \eqref{eq.1914} in \eqref{eq.1915} and using  \eqref{eq.01301423}, \eqref{eq.02011636}, \eqref{eq.1010}, \eqref{eq.102030} and\eqref{eq.01301436}   we have
\begin{align}
&\frac{2}{\pi}\left((E_{\rm bdry}-E_{\rm global})-\Phi_e(r_+)Q_e-\Phi_m(r_+)Q_m\right)=\nn\\
& \frac{2}{\pi}\Bigg( \frac{r_c^3}{L}\left(
-\frac{1}{r_c}\sqrt{1-\frac{\omega}{r_c}+\frac{r_c^2}{L^2}+\frac{q^2}{r_c^2}+\frac{p^2}{r_c^2}-\frac{\alpha'(2B_1+B_2)(q^4+p^4)}{5r_c^6}+\frac{4\alpha' B_1q^2p^2}{5r_c^6}}+\frac{1}{r_c}\sqrt{1+\frac{r_c^2}{L^2}}\right)\nn\\
&-q\left(\frac{q}{r_+}-\frac{2\alpha'(2B_1+B_2)q^3-4\alpha' B_1 q p^2}{5r_+^5}\right)
-p\left(\frac{p}{r_+}-\frac{2\alpha'(2B_1+B_2)p^3-4\alpha' B_1 p q^2}{5r_+^5}\right).
\Bigg)\label{eq.2215}
\end{align}
By equating \eqref{eq.2215} with \eqref{eq.1718}, we find $r_0$ in terms of finite cut off $r_c$.
In the following, we solve this equality in two theories: $F^2$ and $F^4$ theory.
\subsubsection*{Complexity in $ F^2$ dyonic theory at finite cut off}
In this theory, we eliminate $B_1$ and $B_2$ in  eq. \eqref{eq.2215} and \eqref{eq.1718}, and note that for small black holes (i.e. $r_-<r_0<r_+\ll L$) $q^2_e=-q^2_m+r_-r_+$.
By equating \eqref{eq.2215} and \eqref{eq.1718}, at leading order in $r_c$, we have
\begin{align}
r_0=\frac{r_-}{2}\left(1+\frac{q_m^2}{r_- r_+}+\frac{ L^2}{2r^2_c}\left(1+\frac{r_+}{r_-}\right)\right)+\sqrt{\frac{r_-^2}{4}\left(1+\frac{q_m^2}{r_- r_+}+\frac{ L^2}{2r^2_c}\left(1+\frac{r_+}{r_-}\right)\right)^2-q^2_m}.\label{eq.1650}
\end{align}
The above relationship means that the behind horizon cut off $r_0$ is fixed by the UV cut off at the boundary.
If we fixed the UV cut off  $r_c$, we are not allowed to consider another independent cut off inside the horizon, and the UV cut off may probe the degrees of freedoms in black holes interior \cite{Akhavan:2018wla}.

If one substitute the above value of $r_0$ in \eqref{eq.1718} and let $r_c\to \infty$, it is observed that
\begin{align}
\lim_{t\to \infty}\frac{d\CA}{dt}=\frac{q^2_e+q^2_m}{\pi r}\Big\vert^{r_-}_{r_+}.
\end{align}
The above equation, differ by equation \eqref{eq.1707} with a factor $1/2$.
If we tend magnetic charge to zero, the above relation is the  well-known result in \cite{Brown2, Carmi:2017jqz} (which is for electrically charged black holes).

\subsubsection*{Complexity in $ F^4$ dyonic theory at finite cut off}
In $F^4$ dyonic theory, by equating \eqref{eq.2215} and \eqref{eq.1718}, we obtain a  sixth ordered polynomial equation in $r_0$:
\begin{align}
\frac{r_0}{\pi}+\frac{q_m^2}{\pi r_0}+\frac{\alpha'(2B_1+B_2)(q_e^4-q_m^4)}{5\pi r_0^5}=\frac{q_e^2+2q_m^2}{\pi r_+}+\frac{\omega}{2\pi}\frac{L^2}{r_c^2}-\frac{2\alpha'(2B_1+B_2)(q^4_e+2q^2_m)}{5\pi r_+^5}+\frac{12\alpha' B_1 q_e^2 q_m^2}{5\pi r_+^5},
\end{align}
and finding the root of this polynomial is complicated.
However, for small black holes (i.e. $r_0\ll L$), the left hand  side of above equation exactly appear in equation \eqref{eq.1718} and so the  late time growth rate of complexity is (in limit $r_c\to \infty$):
\begin{align}
\lim_{t\to \infty}\frac{d\CA}{dt}=\frac{r_+}{\pi}-\frac{q_e^2+q_m^2}{\pi r_+}+\frac{3\alpha'(2B_1+B_2)(q_e^4+q_m^4)}{5\pi r_+^5}-\frac{12\alpha' B_1 q_e^2 q_m^2}{5\pi r_+^5}.
\end{align}
The above relation is symmetric under transformation $q_e\leftrightarrow q_m$, but apparently it can't be written as \eqref{eq.1707}.

\section{Discussion}
In \cite{Goto:2018iay} , Myers and et al. have investigated complexity for a dyonic black hole, and have found that the late-time growth rate vanishes when the black hole carries only a magnetic charge.
Then they showed that the inclusion of a Maxwell surface term to the action can restore the electric-magnetic duality, and change the value of the late-time growth rate.

In this paper, we have studied the action growth of dyonic black holes with $F^4$ corrections and answered the question of whether in the correction of higher derivatives, these boundary terms are needed.
First, we studied action growth of dyonic black holes in Einstein-Maxwell gravity with quartic field strength corrections in general dimensions, and similar to the four-dimensional   Einstein-Maxwell gravity case, it was concluded that action growth rates vanish for purely magnetic black holes if the matter field boundary terms do not consider, and it violates the electromagnetic duality.

Following the same method of \cite{Goto:2018iay}, and by inclusion  matter field boundary terms (in higher derivatives),  the late time rate of action would become sensitive to the magnetic charge and with the special choice, $\gamma=1/2$,  electric and magnetic charges contribute to action growth on equal footing, which is in accord with electric-magnetic duality.
If we set  $\gamma=1$, then action growth rates vanish for purely electric black holes.

Therefore, it seems that in each order of higher derivative corrections, Maxwell boundary terms (in the same order of derivative)  are needed for the expected behavior of the late time action growth rate, and to satisfy the electric-magnetic duality.

In another way,  we also have studied the growth rate of complexity for $F^2$ and $F^4$ theory with a radial cut off using CA proposal \cite{Akhavan:2018wla} (for small charged black holes, $r_-<r_0<r_+\ll L$).
In this method, Lloyd's bound is a vital principle, this bound relates the late time behavior of complexity which is evaluated from the action behind the horizon to the energy which is charge defined at the UV boundary $r=r_c$.
We have found that if one sets a UV cut off at the boundary, Lloyd's bound enforces us to have a behind horizon cut off whose value is fixed by the UV cut off.
By this consideration, at late times complexity gets linear growth and restores electric-magnetic duality.
In Einstein-Maxwell theory (at $\alpha'^0$ order), the late times complexity is
\begin{align}
\lim_{t\to \infty}\frac{d\CA}{dt}=\frac{q^2_e+q^2_m}{\pi r}\Big\vert^{r_-}_{r_+}, \label{eq.1010.10}
\end{align}
The above answer is twice the value obtained by adding the boundary term, but we do not expect these two methods lead to the same answer.

By UV finite cut off method, complexity growth of dyonic black holes with quartic field strength corrections is also evaluated.
The result obtained by this method is different from the ones obtained with the adding boundary term, but it has the symmetry of electric-magnetic duality

\subsubsection*{Acknowledgment}
I am grateful to  A. Ghodsi for very useful conversation, and valuable guidance  throughout  at various stages of this work. 
I would like to thank S. Qolibikloo, and Gh. Jafari for useful comments.
This work is supported by Ferdowsi University of Mashhad 

\end{document}